\documentstyle[twoside]{article}

\catcode`\@=11
\long\def\@makefntext#1{
\protect\noindent \hbox to 3.2pt {\hskip-.9pt  
$^{{\eightrm\@thefnmark}}$\hfil}#1\hfill}		

\def\@makefnmark{\hbox to 0pt{$^{\@thefnmark}$\hss}}	
	
\def\ps@myheadings{\let\@mkboth\@gobbletwo
\def\@oddhead{\hbox{}
\rightmark\hfil\eightrm\thepage}   
\def\@oddfoot{}\def\@evenhead{\eightrm\thepage\hfil
\leftmark\hbox{}}\def\@evenfoot{}
\def\sectionmark##1{}\def\subsectionmark##1{}}

\oddsidemargin=\evensidemargin
\newcounter{sectionc}\newcounter{subsectionc}\newcounter{subsubsectionc}
\renewcommand{\section}[1] {\vspace{12pt}\addtocounter{sectionc}{1} 
\setcounter{subsectionc}{0}\setcounter{subsubsectionc}{0}\noindent 
	{\tenbf\thesectionc. #1}\par\vspace{5pt}}
\renewcommand{\subsection}[1] {\vspace{12pt}\addtocounter{subsectionc}{1} 
	\setcounter{subsubsectionc}{0}\noindent 
	{\bf\thesectionc.\thesubsectionc. {\kern1pt \bfit #1}}\par\vspace{5pt}}
\renewcommand{\subsubsection}[1] {\vspace{12pt}\addtocounter{subsubsectionc}{1}
	\noindent{\tenrm\thesectionc.\thesubsectionc.\thesubsubsectionc.
	{\kern1pt \tenit #1}}\par\vspace{5pt}}
\newcommand{\nonumsection}[1] {\vspace{12pt}\noindent{\tenbf #1}
	\par\vspace{5pt}}

\newcounter{appendixc}
\newcounter{subappendixc}[appendixc]
\newcounter{subsubappendixc}[subappendixc]
\renewcommand{\thesubappendixc}{\Alph{appendixc}.\arabic{subappendixc}}
\renewcommand{\thesubsubappendixc}
	{\Alph{appendixc}.\arabic{subappendixc}.\arabic{subsubappendixc}}

\renewcommand{\appendix}[1] {\vspace{12pt}
        \refstepcounter{appendixc}
        \setcounter{figure}{0}
        \setcounter{table}{0}
        \setcounter{lemma}{0}
        \setcounter{theorem}{0}
        \setcounter{corollary}{0}
        \setcounter{definition}{0}
        \setcounter{equation}{0}
        \renewcommand{\thefigure}{\Alph{appendixc}.\arabic{figure}}
        \renewcommand{\thetable}{\Alph{appendixc}.\arabic{table}}
        \renewcommand{\theappendixc}{\Alph{appendixc}}
        \renewcommand{\thelemma}{\Alph{appendixc}.\arabic{lemma}}
        \renewcommand{\thetheorem}{\Alph{appendixc}.\arabic{theorem}}
        \renewcommand{\thedefinition}{\Alph{appendixc}.\arabic{definition}}
        \renewcommand{\thecorollary}{\Alph{appendixc}.\arabic{corollary}}
        \renewcommand{\theequation}{\Alph{appendixc}.\arabic{equation}}
        \noindent{\tenbf Appendix \theappendixc #1}\par\vspace{5pt}}
\newcommand{\subappendix}[1] {\vspace{12pt}
        \refstepcounter{subappendixc}
        \noindent{\bf Appendix \thesubappendixc. {\kern1pt \bfit #1}}
	\par\vspace{5pt}}
\newcommand{\subsubappendix}[1] {\vspace{12pt}
        \refstepcounter{subsubappendixc}
        \noindent{\rm Appendix \thesubsubappendixc. {\kern1pt \tenit #1}}
	\par\vspace{5pt}}

\newcommand{\textlineskip}{\baselineskip=13pt}
\newcommand{\smalllineskip}{\baselineskip=10pt}

\def\eightcirc{
\begin{picture}(0,0)
\put(4.4,1.8){\circle{6.5}}
\end{picture}}
\def\eightcopyright{\eightcirc\kern2.7pt\hbox{\eightrm c}} 

\def\abstracts#1#2#3{{
	\centering{\begin{minipage}{4.5in}\baselineskip=10pt\footnotesize
	\parindent=0pt #1\par 
	\parindent=15pt #2\par
	\parindent=15pt #3
	\end{minipage}}\par}} 

\newcommand{\bibit}{\nineit}
\newcommand{\bibbf}{\ninebf}
\renewenvironment{thebibliography}[1]
        {\frenchspacing
	 \ninerm\baselineskip=11pt
         \begin{list}{\arabic{enumi}.}
        {\usecounter{enumi}\setlength{\parsep}{0pt}     
	 \setlength{\leftmargin 12.7pt}{\rightmargin 0pt} 
         \setlength{\itemsep}{0pt} \settowidth
	{\labelwidth}{#1.}\sloppy}}{\end{list}}

\newcounter{itemlistc}
\newcounter{romanlistc}
\newcounter{alphlistc}
\newcounter{arabiclistc}

\newcommand{\fcaption}[1]{
        \refstepcounter{figure}
        \setbox\@tempboxa = \hbox{\footnotesize Fig.~\thefigure. #1}
        \ifdim \wd\@tempboxa > 5in
           {\begin{center}
        \parbox{5in}{\footnotesize\smalllineskip Fig.~\thefigure. #1}
            \end{center}}
        \else
             {\begin{center}
             {\footnotesize Fig.~\thefigure. #1}
              \end{center}}
        \fi}

\newcommand{\tcaption}[1]{
        \refstepcounter{table}
        \setbox\@tempboxa = \hbox{\footnotesize Table~\thetable. #1}
        \ifdim \wd\@tempboxa > 5in
           {\begin{center}
        \parbox{5in}{\footnotesize\smalllineskip Table~\thetable. #1}
            \end{center}}
        \else
             {\begin{center}
             {\footnotesize Table~\thetable. #1}
              \end{center}}
        \fi}

\def\@citex[#1]#2{\if@filesw\immediate\write\@auxout
	{\string\citation{#2}}\fi
\def\@citea{}\@cite{\@for\@citeb:=#2\do
	{\@citea\def\@citea{,}\@ifundefined
	{b@\@citeb}{{\bf ?}\@warning
	{Citation `\@citeb' on page \thepage \space undefined}}
	{\csname b@\@citeb\endcsname}}}{#1}}

\newif\if@cghi
\def\cite{\@cghitrue\@ifnextchar [{\@tempswatrue
	\@citex}{\@tempswafalse\@citex[]}}
\def\citelow{\@cghifalse\@ifnextchar [{\@tempswatrue
	\@citex}{\@tempswafalse\@citex[]}}
\def\@cite#1#2{{$\null^{#1}$\if@tempswa\typeout
	{IJCGA warning: optional citation argument 
	ignored: `#2'} \fi}}

\def\pmb#1{\setbox0=\hbox{#1}
	\kern-.025em\copy0\kern-\wd0
	\kern.05em\copy0\kern-\wd0
	\kern-.025em\raise.0433em\box0}

\def\fnt#1#2{\footnotetext{\kern-.3em
	{$^{\mbox{\scriptsize #1}}$}{#2}}}

\def\fpage#1{\begingroup
\voffset=.3in
\thispagestyle{empty}\begin{table}[b]\centerline{\footnotesize #1}
	\end{table}\endgroup}

\def\runninghead#1#2{\pagestyle{myheadings}
\markboth{{\protect\footnotesize\it{\quad #1}}\hfill}
{\hfill{\protect\footnotesize\it{#2\quad}}}}
\font\tenrm=cmr10
\font\tenit=cmti10 
\font\tenbf=cmbx10
\font\bfit=cmbxti10 at 10pt
\font\ninerm=cmr9
\font\nineit=cmti9
\font\ninebf=cmbx9
\font\eightrm=cmr8

\def\qed{\hbox{${\vcenter{\vbox{			
   \hrule height 0.4pt\hbox{\vrule width 0.4pt height 6pt
   \kern5pt\vrule width 0.4pt}\hrule height 0.4pt}}}$}}

\def\bsc{{\sc a\kern-6.4pt\sc a\kern-6.4pt\sc a}}  	
\def\bflatex{\bf L\kern-.30em\raise.3ex\hbox{\bsc}\kern-.14em 
T\kern-.1667em\lower.7ex\hbox{E}\kern-.125em X} 

\begin{document}

\runninghead{A New Look at the 2D Ising Model}
            {A New Look at the 2D Ising Model}

\normalsize\textlineskip
\thispagestyle{empty}
\setcounter{page}{1}
\fpage{1}
\centerline{\bf A NEW LOOK AT THE 2D ISING MODEL FROM EXACT}
\vspace*{0.035truein}
\centerline{\bf PARTITION FUNCTION ZEROS FOR LARGE LATTICE SIZES}
\vspace*{0.37truein}
\centerline{\footnotesize NELSON A. ALVES\footnote{E-mail: 
             alves@fma.if.usp.br}
                \, and J. R. DRUGOWICH DE FELICIO\footnote{E-mail: 
             drugo@pinguim.ffclrp.usp.br}}
\vspace*{0.015truein}
\centerline{\footnotesize\it Departamento de F\'{\i}sica e Matem\'atica, FFCLRP,
    Universidade de S\~ao Paulo}
\baselineskip=10pt
\centerline{\footnotesize\it Av. Bandeirantes 3900, CEP 014040-901 Ribeir\~ao Preto, SP, Brazil}
\vspace*{10pt}
\centerline{\footnotesize ULRICH H. E. HANSMANN\footnote{E-mail: 
hansmann@ims.ac.jp}}
\vspace*{0.015truein}
\centerline{\footnotesize\it Department of Theoretical Studies, Institute for Molecular Science}
\baselineskip=10pt
\centerline{\footnotesize\it Okazaki, Aichi 444, Japan}
\vspace*{0.225truein}
\abstracts{A general numerical method is presented to locate the
partition function zeros in the complex $\beta$ plane for large
lattice sizes.  We apply this method to the 2D Ising model and
results are reported for square lattice sizes up to $L=64$.
We also propose an alternative method to evaluate corrections
to scaling which relies only on the leading zeros. This method
is illustrated with our data.}{}{}

\vspace*{10pt}


\vspace*{1pt}\textlineskip	
\section{Introduction}   	
\vspace*{-0.5pt}
\noindent
   Although the two-dimensional Ising model is exactly solved
for zero external field, it continues receiving attention in many
aspects.
   In a recent work by Beale$^1$ for instance, the low-temperature
series expansion for the partition function was exactly determined 
for finite lattices with periodic boundary conditions. 
   In terms of the expansion variable $ u= e^{-4 \beta}$ the partition 
function on a $m\,{\rm x}\, n$ lattice size becomes a polynomial of 
finite degree in $u$, and its coefficients $g(E)$, the number of 
configurations with energy $E$, were calculated from an exact closed 
form based on Kaufman's solution.$^2$

    Motivated by that calculation as well by the enhancement of computer
facilities we decided to revisit the 2D Ising model to obtain the 
exact partition function zeros in the complex temperature plane.$^3$
    This approach had already been pursued by Katsura and Abe$^{4,5}$  
in the early investigations of zero distributions 
in order to check the proposal by Fisher about their loci.$^6$
    Other papers have aimed the study of the critical properties  
from the leading zeros, but all of them were limited to 
small lattices ($m\,{\rm x}\, n \le 13\,{\rm x}\, 13$).$^{7,8,9}$

   In this work we present a procedure to obtain exact complex zeros
of the partition function for large lattice sizes.  
   We provide a description of a way round technical limitations
on solving polynomials, at least in what concerns the location of 
the first zeros  ($u_1^0$, $u_2^0$, ...).
   It is a modified  version of the scanning procedure which has been 
applied to continuous energy distributions of lattice gauge 
theories.$^{10,11,12}$
   Our approach is based on constructing a new function
in terms of $~{\rm ln}\,g(E)$. It is presented in section 2, where the
first exact zeros are calculated for square lattice sizes up to $L = 64$.
   Since we are working 
with numerical computation, we mean by exact zeros accurate values 
limited only by the use of double-precision floating-point arithmetic.

   The partition function zeros approach has been largely used to obtain 
information on phase transitions from  
Monte Carlo (MC) simulations$^{3,10,13}$ or exact enumerations$^{9}$
of finite systems.
   In this context, Itzykson's {\it et al.}$^3$ finite size scaling (FSS) relation for the 
first zero is the proper way to calculate the correlation exponent $\nu$.
   We present results so obtained in section 3. 
   In addition, looking forward to obtain more information from those
exact zeros, we propose, in the same section, a new way to evaluate corrections to FSS
which relies only on $u_1^0(L)$ data.

\section{Exact partition function zeros}
\noindent
    The partition function of the two-dimensional Ising model 
on a $m\,{\rm x}\, n$ lattice can be written as a polynomial, 
\begin{equation}
 Z_{n,m}(\beta) = e^{2nm\beta} \sum_{E=0}^{nm} g(E) u^E \label{eq:z1}
\end{equation}
where $u = e^{-4\beta}$.

Kaufman's solution for the isotropic Ising model renders the 
analytical expressions to be expressed in the polynomial form (\ref{eq:z1}).
   Following Beale$^1$, this can be done for any lattice
size by using MATHEMATICA.

   A further step namely exact determination of their zeros and FSS analysis 
for the leading ones,  can be achieved with this polynomial form. However we
have checked that it was not possible to handle systems for $L$ larger than $16$ 
with our workstations. In fact, as the lattice size increases, the exact coefficients
become very large integers. The enormous increasing of their maximum values, 
typically ${\rm ln}\, g(E) \simeq 174$ for $L=16$ and going up to
${\rm ln}\, g(E) \simeq 2835$ for $L=64$, prevent us from  solving them 
by using computer algebra language. The same reasons do not indicate the use 
of standard numerical algorithms,$^{14,15}$ usually 
employed in those cases. 

   To circumvent this problem we borrowed inspiration from lattice
gauge theories where the energy distributions are continuous.
   There, in contrast to spin systems where the action
takes discrete values and the partition function becomes
a polynomial in $u$, a time series analysis in function of the 
complex coupling $\beta = \beta_x + i \beta_y$ 
is more efficient in calculating the first zero 
which is closest to the infinite volume critical point.
   It is a two step approach.$^{12}$ 
   First, we scan graphically for separate zeros of 
${\rm Re}\, Z(\beta)$ and ${\rm Im}\, Z(\beta)$,
where
\begin{equation}
 Z(\beta) = \sum_E g(E) u^{\triangle E}\,. \label{eq:z2}
\end{equation}
    The shift $\,\triangle E\,=\,E~-<E>\,$ is usually introduced for
technical reasons, to avoid numerical overflow computations, 
although it is not any more relevant in the new approach.
    A typical output is shown in Fig.\,1. 
    Crosses correspond to ${\mbox{\rm Re}\, Z(\beta) = 0}$ and diamonds 
to ${\mbox{\rm Im}\, Z(\beta) = 0}$.
    The wanted zero is obtained when the lines cross.
    Second, we compute this zero to a desired precision as an
iterative process.
    This can be achieved by means of the minimization algorithm AMOEBA$^{16}$ 
for the function 
$(({\rm Re}\, Z(\beta)) ^2\,+\,({\rm Im}\, Z(\beta)) ^2)^{1/2}$,  
whose starting point is obtained from a simple
inspection of figures like Fig. 1.
    As an example, we can use the input (0.43765, 0.0131)
as the starting point to this routine
which leads, after roughly 100 iterations, to $(\beta_x^0, \, \beta_y^0) = 
(0.4376431265, ~0.01311604331)$.

    Now we shall describe how to implement our approach.

    Since our aim is to achieve large lattices one has to work with 
logarithms.
    For this end we need to introduce a new function 
$ F(\beta) = F_x(\beta) + i F_y(\beta)$ to play the role of  $Z(\beta)$
itself.
    We start from 
\begin{figure}[hb]
\begin{center}
\input{f1.inp}
\fcaption{Search for the first partition function zero for $L=64$. The crosses
        indicate the zeros for ${\rm Re}\,F(\beta)$ and the diamonds the 
        ones for ${\rm Im}\,F(\beta)$. The complex function $F(\beta)$ has the same zeros
        as $Z(\beta)$.}
\end{center}
\end{figure}
splitting ${\rm Re}\,Z(\beta)$ into two positive parts, namely $G_E$ and $H_E$,
defined by
\begin{equation}
   {\rm Re}\,Z(\beta) = G_E - H_E
\end{equation}
where
\begin{equation}
   G_E = \sum_{E} \raisebox{.6ex}{$'$}~
       g(E)~ e^{-4\beta_x \triangle E}~ {\rm cos}(\alpha_E)\,, \label{eq:g}
\end{equation}
\begin{equation}
    H_E = \sum_{E} \raisebox{.6ex}{$''$} ~
       g(E)~ e^{-4\beta_x \triangle E}~ |{\rm cos}(\alpha_E)|\,, \label{eq:h}
\end{equation}
and  $ \alpha_E = 4\beta_y \triangle E $.
    Here $\sum \raisebox{.3ex}{$'$}\,$ means a summation over $E$ provided ${\mbox {\rm cos}(\alpha_E)\,>\,0}$, 
 and $\, \sum \raisebox{.3ex}{$''$}\,$ stands for the complementary values where ${\rm cos}(\alpha_E)\,<\,0$.
    Next we calculate  ${\rm ln}\, G_E$ and ${\rm ln}\, H_E$ in a recursive way from the terms 
$ {\rm ln}\, g(E)\,$, ${\rm ln\, cos}(\alpha_E)$   
$\,{\mbox{\rm and }~  -4\beta_x \triangle E}$.
    Logarithmic terms can be added up two by two in $G_E$ and $H_E$,
respectively, by using the relation 
\begin{equation}
{\rm ln} (a + b) = {\rm ln}\,b +
         {\rm ln} ( 1 + e^{ {\rm ln}\, a - {\rm ln}\, b} )\,. \label{eq:lnab}
\end{equation}
    Finally, since we are interested in the $\beta$ values 
where  ${\rm Re}\, Z(\beta)$  and ${\rm Im}\, Z(\beta)$ change signals, 
we realize that this can be achieved by the function
\begin{equation}
F_x(\beta) =  {\rm ~ln}~G_E - {\rm ~ln}~H_E \,,
\end{equation}
and a similar one for $F_y(\beta)$, which follows from 
the imaginary part of $\,Z(\beta)$.
 
    Now, we apply the two steps procedure to find roots of
$ F_x(\beta) ~{\rm and}~ F_y(\beta)$ instead of ${\rm Re}\,Z(\beta)$ and
${\rm Im}\,Z(\beta)$.
   In Fig.\,1 we show the first step for $L=64$. 
   Table 1 contains our leading zeros $u_1^0(L)$ for lattice sizes up to
$L = 64$, with rounded errors in the last digit.
\begin{table}[htbp]
\tcaption{First partition function zeros.}
\centerline{\footnotesize\smalllineskip
\begin{tabular}{l c c c c}\\
\hline
~~$L$ &{} &~~~~~Re$(u_1^0)$ &{} &~~~~~~Im$(u_1^0)$\\
\hline
~~4 & & ~~$0.1624473772$ & & ~~$0.16648190032$ \\  
~~6 & & ~~$0.1756913616$ & & ~~$0.10528348725$ \\
~~8 & & ~~$0.1780809275$ & & ~~$0.07710375572$ \\
~~9 & & ~~$0.1783370200$ & & ~~$0.06801661701$ \\
~10 & & ~~$0.1783571854$ & & ~~$0.06084948478$ \\
~12 & & ~~$0.1780873239$ & & ~~$0.05026266796$ \\
~15 & & ~~$0.1774653671$ & & ~~$0.03986421638$ \\
~16 & & ~~$0.1772557409$ & & ~~$0.03729300267$ \\ 
~18 & & ~~$0.1768587209$ & & ~~$0.03303236187$ \\
~20 & & ~~$0.1764984476$ & & ~~$0.02964570468$ \\
~24 & & ~~$0.1758873488$ & & ~~$0.02460155919$ \\ 
~30 & & ~~$0.1751918649$ & & ~~$0.01959967784$ \\
~32 & & ~~$0.1750048586$ & & ~~$0.01835571819$ \\ 
~36 & & ~~$0.1746815338$ & & ~~$0.01628818111$ \\
~40 & & ~~$0.1744124041$ & & ~~$0.01463927731$ \\
~48 & & ~~$0.1739912845$ & & ~~$0.01217440355$ \\
~60 & & ~~$0.1735492820$ & & ~~$0.00971963267$ \\
~64 & & ~~$0.1734355215$ & & ~~$0.00910750889$ \\ 
~$\infty$ &&~~$0.1715728753$ & & \\
\hline\\
\end{tabular}}
\end{table}

\begin{figure}[htb]
\begin{center}
\input{f2.inp}
\fcaption{The complete set of partition function zeros for $L=16$ in the
        complex $u$ plane (diamonds). The continuous line,
        parametrized by Re$(u) = 1 + 2^{3/2}{\rm cos}\,w\,+\,2\,{\rm cos}\,2w$
        and
        Im$(u) = 2^{3/2}{\rm sin}\,w\,+\,2\,{\rm sin}\,2w$, for $0 \leq w < 2 \pi$
        corresponds to the phase boundaries (Ref.\,17).} 
\end{center}
\end{figure}

   This method can  easily be implemented and takes 
few minutes of CPU time for a Fortran code in a workstation
after we have calculated the coefficients $g(E)$.$^1$
   The lattice sizes were chosen to explore finite size corrections
by an alternative method which will be presented  in the next section.
   In addition, we plot in Fig. 2 all zeros for $L=16$ 
obtained with MATHEMATICA and compare them with the
expected phase boundaries in the $u$ plane for 
$L \rightarrow \infty$.$^{17,18}$ 
  This curve corresponds to the locus of points 
where the free energy is non-analytic and it is parametrized by
Re$(u) = 1 + 2^{3/2}\,{\rm cos}\,w \,+\,2\,{\rm cos}\,2w\,$ and
Im$(u) = 2^{3/2}\,{\rm sin}\,w\,+\,2\,{\rm sin}\,2w$, for $~0 \leq w < 2 \pi$. 

\section{Finite size scaling analysis}
\noindent
  The systematic dependence of $\nu$ on finite systems
can be explored to evaluate the main correction to scaling.
  From pairs of lattices $L$ and $L'$, we define the corresponding 
finite size estimators, 
\begin{equation}
\frac{1}{\nu_{L,L'}} \, = \, {\rm ln} \left( \frac{|u_1^0(L')-u_c|}
       {|u_1^0(L)-u_c|} \right) / \,{\rm ln} (\frac{L}{L'})\,. \label{eq:nu}
\end{equation}
   This equation was already used to estimate the critical
exponent $\nu$ for the 3D Ising model from increasing pairs of lattices.$^{19}$

\begin{table}[htbp]
\tcaption{Sequence of estimates for $1/ \nu_{L,sL}$.}
\centerline{\footnotesize\smalllineskip
\begin{tabular}{l c c c c}\\
\hline
~$L~~/~s$ &~~~~1.5    &~~~~~2      &~~~~~3      &~~~~~4 \\ 
\hline
~4       &1.131948204 &1.107535391 &1.083907993 &1.071996844 \\    
~6       &1.067288812 &1.055806270 &1.044309560 &1.038354523 \\
~8       &1.043516688 &1.036458296 &1.029248898 &1.025453193 \\
10       &1.031641084 &1.026690577 &1.021565572 &1.018835580 \\
12       &1.024655805 &1.020902776 &1.016977082 &1.014867660 \\   
16       &1.016924350 &1.014448089 &1.011818989 &1.010388613 \\    
20       &1.012804317 &1.010980582 &1.009024731 &            \\
24       &1.010266064 &1.008832544 &            &            \\
32       &1.007324514 &1.006329138 &            &            \\   
40       &1.005681181 &            &            &            \\
\hline\\
\end{tabular}}
\end{table}

   In Table 2 we present sequences of $1/\nu_{L,sL}$ as a function
of a fixed rescaling factor $ s = L'/L$.
   As min$(L,L')$ increases, the values obtained by matching
pairs of lattices approach the expected limiting value  $\nu = 1$.

   Equation (\ref{eq:nu}), beyond being quite important to estimate $\nu$
can be a starting point to evaluate finite size corrections, which can be 
due to a variety of sources.$^{20,21}$
   For this end let us briefly recall Nightingale's finite size RG 
transformation.

   Under the hypothesis the system is large enough to consider
the scaling relation for the longitudinal correlation length $\xi_L(\beta)$,
the standard expression for the correlation exponent $\nu$ is$^{22,23}$
\begin{equation}
 1+ \frac{1}{\nu_{L,L'}} \,  = \,
              {\rm ln}   \left( \frac{
               \rm{\raisebox{.4ex}{$\partial \xi_{L'}$}} /\partial\beta}
             { \rm{\raisebox{.4ex}{$\partial \xi_{L}$}} / \partial\beta}
               \right)_{\!\! \beta_c}  
             / \, {\rm ln} (\frac{L'}{L})\,,    \label{eq:brezin}
\end{equation}
  where the scaling equation for the correlation length is given by 
\begin{equation}
\xi_L \,=\, L\, Y_{\xi}(\,(\beta-\beta_c)L^{1/ \nu}, hL^{y_H}, uL^{y_3})\,.
                                                \label{eq:night}
\end{equation}
   This differentiable equation includes corrections due to the 
leading bulk irrelevant scaling field $u$ with exponent $y_3 < 0$,
and a magnetic field dependence for the sake of completeness.

   From Eq.\,(\ref{eq:brezin}) and Eq.\,(\ref{eq:night}) one
obtains, for $h=0$,
\begin{equation}
 \frac{1}{\nu_{L,L'}} \, =\, \frac{1}{\nu}            \, 
  +\,  a_0 \frac{L'^{y_3} - L^{y_3}}{{\rm ln}(L'/L)}  \,
  +\,  b_0 \frac{L'^{2 y_3} - L^{2 y_3}}{{\rm ln}(L'/L)} \,
  +\, ...                    \label{eq:cross}
\end{equation}
where
$a_0$ and $b_0$ include derivatives like  
${\partial Y_{\xi}(y,z)} / {\partial y} |_{y=0,z=0} $.
   If we replace $L \rightarrow L'-1$ in the above
equation, one obtains Privman and Fisher's results.$^{23}$

   Our analysis follows with the introduction of the 
rescaling factor $s$ in Eq.\,(\ref{eq:cross}).
   The exponent $\nu$ can be asymptotically obtained from sequences
either by extrapolating ${\rm ln}\,s \rightarrow \infty$
or by extrapolating  ${\rm ln}\,s \rightarrow 0$ (see the similar scaling behaviour of 
Binder's function$^{24}$ $W_{\beta}^*$ for obtaining $2 \beta / \nu$).
   The exponent $y_3$ can be evaluated through a linear regression 
for finite lattice sizes $L$ if we fix the ratio $s$.
   At this point we call attention to the known fact that the
main finite size correction for the 2D Ising model comes from 
nonlinear scaling fields$^{23,25}$ and
gives origin to different $L-$dependent corrections in
Eq.\,(\ref{eq:cross}).

\begin{table}[htbp]
\tcaption{Estimates of $w$ from $1/ \nu_{L,sL}$ data in Table 2. 
         $\tilde{w}$ refers to 
         $1/ \nu_{L,sL}$ data obtained with the replacement
         $|u_1^0(sL)-u_c|\,/ \,|u_1^0(L)-u_c|$  
      by ${\rm Im}\,u_1^0(sL)\,/ \, {\rm Im}\,u_1^0(L)$ 
          in Eq.~8. }
\centerline{\footnotesize\smalllineskip
\begin{tabular}{l c c c c c c c}\\
\hline
~$s$& ~~~~~~~~$L$          &~&~~$-w$     &~${\rm- ln}\,a(s)$  &~~& 
                    ~~$-\tilde{w}$ &~${\rm- ln}\,\tilde{a}(s)$\\ 
\hline
1.5 &~(4,6,8,10,12,16,20,24,32,40)& &1.34903  &~~0.284605 & &  
                                     1.09706  &~~0.586739 \\    
1.5 &~(12,16,20)                  & &1.28388  &~~0.514528 & &
                                     1.05752  &~~0.716982 \\    
1.5 &~(12,16,20,24,32,40)         & &1.21665  &~~0.698876 & &    
                                     1.03807  &~~0.770224 \\    
1.5 &~(20,24,32,40)               & &1.17201  &~~0.850741 & &        
                                     1.02581  &~~0.811914 \\    
1.5 &~(32,40)                     & &1.13858  &~~0.970497 & &   
                                     1.01865  &~~0.837543 \\
{}  & {} & {} & & & & & \\
2   &~(4,6,8,10,12,16,20,24,32)   & &1.35083  &~~0.458201 & &   
                                     1.09585  &~~0.735457 \\
2   &~(12,16,20)                  & &1.26133  &~~0.735541 & &
                                     1.05021  &~~0.873225 \\
2   &~(12,16,20,24,32)            & &1.21780  &~~0.853411 & &    
                                     1.03808  &~~0.906048 \\   
2   &~(20,24,32)                  & &1.17102  &~~1.00516  & &    
                                     1.02555  &~~0.946659 \\      
{}  & {} & {} & & & & & \\
3   &~(4,6,8,10,12,16,20)         & &1.37931  &~~0.624774 & &     
                                     1.10630  &~~0.903252 \\     
3   &~(12,16,20)                  & &1.23804  &~~1.00130  & &     
                                     1.04348  &~~1.06970  \\    
{}  & {} & {} & & & & & \\
4   &~(4,6,8,10,12,16)            & &1.39591  &~~0.736397 & &  
                                     1.11314  &~~1.01358  \\    
4   &~(12,16)                     & &1.24609  &~~1.11215  & & 
                                     1.04606  &~~1.18084  \\
\hline\\
\end{tabular}}
\end{table}

   Let us call $w$ the effective exponent coming from the equation
\begin{equation}
{\rm ln}\, \left( \frac{1}{\nu_{L,sL}} - \frac{1}{\nu} \right) =
             w\,{\rm ln}\,L + {\rm ln}\,a(s)\,,  \label{eq:w}       
\end{equation}
which intends to detect the main correction regardless its origin.

   We collect in Table 3 our results for $w$.
   The third and fourth columns correspond to fit Eq.~(\ref{eq:w}) 
to data of Table 2.

   For the 2D Ising model $u_c$ is exactly known, however for many
models the value of $u_c$ is not known with high precision and 
in this case it is usual to replace $|u_1^0 - u_c|$ by Im $u_1^0$. 
   For sake of illustrative purposes both cases were considered.
   In our tables we use the notation $\tilde{w}$ instead of $w$ when
the $1/ \nu_{L,sL}$ data is obtained from Eq.\,(\ref{eq:nu}) 
with the replacement
         $|u_1^0(sL)-u_c|\,/ \,|u_1^0(L)-u_c|$ 
      by ${\rm Im}\,u_1^0(sL)\,/ \, {\rm Im}\,u_1^0(L)$. 
   Different fixed ratios $s$ are used to show the behaviour
of $w$.
   As $s$ increases, and consequently $L'$, the numerical results show 
a trend in direction of $w\,=\,-1$.
   Large $s$ means working with crossings involving a large $L$,
hence close to $\beta_c$. 
   The corresponding best linear fits for all data 
are presented in Fig. 3.
   It is clear that small lattice sizes give origin to deviations
in the employed linear equation (\ref{eq:w}).

   To complete our analysis we present in Table 4 the 
dependence of $w$ on small lattices.
   The value $w\,\approx\,- 1.7$ is quite close to Binder's reported 
value, $w\,\approx\,- 1.8$, as the main correction to the function
$W_{\beta}^*$.$^{24,26}$
   However there smaller lattices were considered in the analysis
which seem to increase $|w|$.
   This trend can be caught from our smaller data set with $s=1.5$ 
for $L = 4 \,{\rm and}\, 6$.

\begin{figure}[htb]
\begin{center}
\input{f3.inp}
\fcaption{Linear regressions for ${\rm ln}\,(1/\nu_{L,sL}\,-1)$ for several values of 
          the rescaling factor $s$, according to Eq.\,(12).}
\end{center}
\end{figure}

  In summary, we have described an approach to compute partition function
zeros for large lattice sizes.
  Although it was applied to the exact 2D Ising model partition 
function, it can be quite useful either when we deal with large 
coefficients, even out of scope of double precision computations,
or when the polynomial has a 
large number of coefficients, which would prevent us from using  
standard solving algorithms. This last situation is proper
for the multicanonical simulation of the density of states
in the 3D Ising model.$^{27}$
 
    In MC simulations FSS behaviour of the first zero has been 
used to estimate the exponent $\nu$ and the critical coupling $\beta_c$.
    Beyond any MC data, with limited statistical precision, we were 
able to explore here the performance of the FSS analysis proposed
to evaluate corrections to scaling.
\begin{table}[hb]
\tcaption{Estimates of $w$ as in Table 3. }
\centerline{\footnotesize\smalllineskip
\begin{tabular}{l c c c c c c c}\\
\hline
~$s$& ~~~~$L$     &~&~~$-w$     &~${\rm- ln}\,a(s)$  &~~& 
                    ~~$-\tilde{w}$ &~${\rm- ln}\,\tilde{a}(s)$\\ 
\hline 
1.5 &~(4,6)       & &1.66085  &~-0.277077               & & 1.27701  &~~0.268877 \\    
1.5 &~(4,6,8)     & &1.60430  &~-0.191966               & & 1.23731  &~~0.328639 \\    
1.5 &~(4,6,8,10)  & &1.56144  &~-0.123144               & & 1.20877  &~~0.374454 \\    
{} & & & & & & & \\
2   &~(4,6)       & &1.61773  &~-1.27172$\,{\rm x}\,10^{-2}$& & 1.24410  &~~0.478115 \\
2   &~(4,6,8)     & &1.56425  &~~6.77855$\,{\rm x}\,10^{-2}$& & 1.20847  &~~0.531757 \\
2   &~(4,6,8,10)  & &1.52369  &~~0.132909               & & 1.18311  &~~0.572470 \\   
\hline\\
\end{tabular}}
\end{table}
    This approach reveals to be quite useful in setting an upper limit on $w$
for the same data set as $s$ increases, as can be observed, for example, from 
our Table 3 for $L=12, 16$ and $20$.

\nonumsection{Acknowledgements}
\noindent
   Nelson Alves and Drugowich de Felicio are supported by the Brazilian
agencies CNPq and FAPESP. 
    Numerical calculations were performed in part on the
Alpha-AXP 3000/300X station at FFCLRP and in part on the
Alpha-AXP 3000/800 at Departamento de F\'{\i}sica Matem\'atica
(IFUSP) in S\~ao Paulo.

\nonumsection{References}

\end{document}